\documentclass[preprint,aps,showpacs,amsmath,amssymb]{revtex4}

\usepackage{graphicx}
\usepackage{dcolumn}
\usepackage{bm}
\usepackage[centerlast]{caption2}

\begin{document}

\title{Empirical comparison of high gradient achievement for different metals in
DC and pulsed mode}

\author{F. Le Pimpec}
 \email{frederic.le.pimpec@psi.ch}
\author{C.~Gough}%
\author{M.~Paraliev}%
\author{R.~Ganter}%
\author{C.~Hauri}%
\author{S.~Ivkovic}%

\affiliation{Paul Scherrer Institute, CH-5232 Villigen PSI,
Switzerland}

\date{\today}

\begin{abstract}
For the SwissFEL project, an advanced high gradient low emittance
gun is under development. Reliable operation with an electric
field, preferably above 125~MV/m at a 4~mm gap, in the presence of
an UV laser beam, has to be achieved in a diode configuration in
order to minimize the emittance dilution due to space charge
effects. In the first phase, a DC breakdown test stand was used to
test different metals with different preparation methods at
voltages up to 100~kV. In addition high gradient stability tests
were also carried out over several days in order to prove reliable
spark-free operation with a minimum dark current. In the second
phase, electrodes with selected materials were installed in the
250~ns FWHM, 500~kV electron gun and tested for high gradient
breakdown and for quantum efficiency using an ultra-violet laser.
\end{abstract}

\maketitle

\section{Introduction}

The Paul Scherrer Institut (PSI) is interested in constructing and
operating a 4$^\mathrm{th}$ generation light source before the end
of the next decade. This X-Ray Free Electron Laser (XFEL),
operating at $\lambda$ = 1~\AA, will be a companion to the
existing 3$^\mathrm{rd}$ generation light source, the Swiss Light
Source. The SwissFEL (previously PSI-XFEL) facility, in order to
fit the site and the budget constraint, needs a concept which
reduces the required beam energy given by the desired radiation
wavelength $\lambda$ according to the spatial coherence criterion,
see equation~\ref{eq:one}

\begin{equation}
\frac{\varepsilon^{(n)}}{\beta \gamma} < \frac{\lambda}{4\pi},
\label{eq:one}
\end{equation}

\noindent where $\varepsilon^{(n)}$ is the normalized transverse
emittance at the undulator entrance, and $\beta \gamma$ being the
relativistic coefficient \cite{Leeman:2007, Oppelt:FEL07}.

In order to produce a low emittance beam, studies on a prototype
diode electron gun have been performed. The diode should sustain
the highest possible gradient in order to accelerate a high charge
density beam and minimize emittance dilution due to space charge
effects. The goal is to exceed typical values of accelerating
gradient achieved in state of the art photoguns ($\sim$ 100~MV/m).
As discussed in \cite{lepimpec:2007}, reviewing over a century of
research on vacuum breakdown is not sufficient to select the
appropriate material for a specific application.

Metallic electrodes, with different preparation procedures were
first tested in a 100~kV DC system, and the results obtained
compared to the literature. Parts of those results were already
presented in \cite{lepimpec:2007}. As the diode prototype came on
line high gradient (HG) high voltage (HV) tests were carried out
with and without the UV laser. The results, including quantum
efficiency (QE) measurements, are reported in this paper.

\section{100 kV DC System}

\begin{figure}[htb]
   \centering
   \includegraphics*[width=0.8\columnwidth]{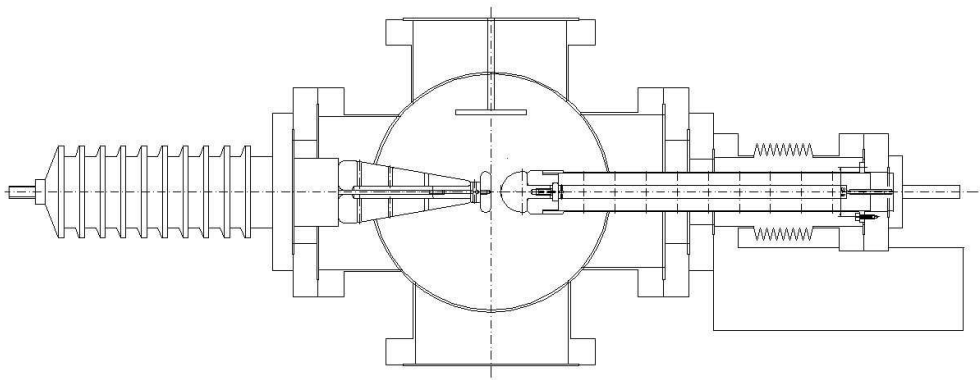}
   \caption{Diode configuration of the 100~kV DC chamber. The cathode is flat the anode is hemispheric
    and the third flat electrode above is used for plasma glow discharge.}
   \label{figTestStand}
\end{figure}

The ultra-high vacuum (UHV) chamber shown in
Fig.\ref{figTestStand}, is described in detail in
\cite{lepimpec:2007}. The UHV system comprises a diode ion pump of
150~l/s, an injection line, a leak valve and a capacitance
pressure gauge. The gauge measures the absolute pressure of the
injected gas (Ar, He or a mixture of both) in order to prepare for
plasma glow discharge (PGD) between the electrodes ($\sim$ 20
minutes, 500~eV, pressure.distance $<$ 1~torr.cm). Polished
electrodes were treated using only Argon gas, providing ions of
400~eV (ArPGD). The ion current was $\sim$10~mA spread over a few
cm$^2$ of electrode. PGD for the rough electrodes were, at first,
a mixture of He-Ar followed by pure Ar \cite{lepimpec:2007}.

A negative DC bias of 0 to 100~kV is applied to the cathode
through an insulating ceramic, (on the left in
Fig.\ref{figTestStand}). The anode is grounded. In principle the
power supply can deliver up to one mA at 100~kV. It is possible to
adjust the maximum current allowed to be delivered by the power
supply; we have adjusted it around one microA. The potential
energy stored in the system is 1.5~J at 100~kV. The current
flowing from the cathode to the anode is measured across a
1~M$\Omega$ resistor with a Keithley multimeter and recorded with
Labview{\texttrademark} software. The gap separation between the
electrodes is adjustable with a translation feedthrough and is
measured with a digital micrometer. A photomultiplier (PMT) is
used to detect X-ray (XR) activity.

Usually one pair of electrodes is tested. They are then
re-machined and polished for subsequent tests. In the cases of
copper, aluminium and stainless steel two different pairs have
been used.

\subsection{Non Polished Electrodes}

The modus operandi and a first set of results were published in
\cite{lepimpec:2007} and is summarized in Table
\ref{tabDCunpolished}. The electrodes were, for the most part,
rough machined and their average surface roughness (R$_a$) is
below, but close to, 200~nm. The highest gradients were usually
obtained after the first in-situ PGD cleaning but sometimes after
a second. The values quoted for 1~nA were, in general, obtained
after arcing. After arcing, field emission (FE), or "dark
current", appears for lower gradients. A subsequent PGD can often
restore the gradient without FE.

\begin{table}[htbp]
\centering

\caption{DC gradient in MV/m before breakdown between cathode
(Cthd) - anode (And) at 1~mm gap (unless specified otherwise) for
the given dark current in nA.}\

\begin{tabular}{|c|c|c|c|}
\hline Cthd-And& Dark Current  &  As Received & After PGD\\
\hline \raisebox{-1.50ex}[0cm][0cm]{SS-SS}& $<\  0.05$ nA & 40& 68 \\
\cline{2-4} &1 nA & 42.5 & 35 \\
\hline \raisebox{-1.50ex}[0cm][0cm]{Ti-Ti}& $<\  0.05$ nA& 50& 63\\
\cline{2-4} & 1 nA& 46.6 & 67 (0.1nA) \\
\hline \raisebox{-1.30ex}[0cm][0.2cm]{Ti-Ti}& $<\  0.05$ nA &  29.6 & 39  \\
\cline{2-4} Vac Fired & 1 nA & 32.5 & 41.4 \\
\hline \raisebox{-1.30ex}[0cm][0.2cm]{Mo-Mo}& $<\  0.05$ nA& 37 & 44 \\
\cline{2-4} Vac Fired & 1 nA &  45.2& 61.3 \\
\hline \raisebox{-1.3ex}[0cm][0.2cm]{Cu-Cu}& $<\  0.05$ nA& - & 32 \\
\cline{2-4} Oxidized & 1 nA & - & 29.3 \\
\hline \raisebox{-1.3ex}[0cm][0.2cm]{Cu-Cu}& $<\  0.05$ nA& 24 & 55 \\
\cline{2-4} Etched & 1 nA &  26 & 19 \\
\hline \raisebox{-1.50ex}[0cm][0cm]{Cu-Mo}& $<\  0.05$ nA& 18.2 (3mm) & 21.6\\
\cline{2-4}  & 1 nA &  13.8 (3mm)& 25.4 \\
\hline \raisebox{-1.50ex}[0cm][0cm]{Al-Al}& $<\  0.05$ nA& - & 52 \\
\cline{2-4}  & 1 nA & 7.5 & 30 \\
\hline \raisebox{-1.0ex}[0cm][0.0cm]{Al-Al}& $<\  0.05$ nA& 36 (2mm) & 73\\
\cline{2-4} (*) & 1 nA &  29 & 31 \\
\hline \raisebox{-1.50ex}[0cm][0cm]{Nb-Nb}& $<\  0.05$ nA& - & 10 (4mm)\\
\cline{2-4}  & 1 nA & - & 5.5 (4mm) \\
\hline
\end{tabular}
\begin{tabular}{c}
SS: Stainless steel ; (*) Mirror-finished - damaged anode
\end{tabular}
\label{tabDCunpolished}
\end{table}

The PGD between the electrodes is obtained by biasing the upper,
third, electrode positively up to 500~V and connecting the
anode-cathode pair to ground \cite{lepimpec:2007}. This
configuration cleans the electrode pair in a similar manner.
Applying the bias to each electrode, one after the other,
frequently results in transferring contaminants from one electrode
to the other. However, in DC mode a breakdown is generally cathode
initiated \cite{springer5}. Therefore, cleaning the cathode last
can still be beneficial in reaching high gradient.

Most of the metals tested in this first sample set reached their
breakdown gradient without arcing. Ti and Mo sustained a few
micro-breakdowns before arcing. Micro-breakdown in this system is
characterized by a surge in pressure, followed by a burst (a few
seconds) of FE current. A definitive breakdown is defined and
characterized by a sustained vacuum activity. The pressure is
often ten times higher than during normal operation, and the FE
current is such that the 100~kV power supply cannot hold the
voltage. Electric gradient values are then recorded when the FE
current is set to 1~nA, to compare with the literature
\cite{Furuta:Nima2005}. This explains why values at 1~nA are
generally lower than the values without FE ($<$0.05~nA), as in the
case of the etched Cu after PGD, Table.\ref{tabDCunpolished}

\subsection{Polished Electrodes}

Electrodes were polished or diamond turned by commercial companies
capable of producing high quality optical mirrors. The average
roughness, R$_a$, of the polished electrodes was below 40~nm, and
below 10~nm for Al, Cu and SS. The term "mirror-finished" will
refer to surfaces with average roughness below 40~nm. The
breakdown gradients obtained were compared to those obtained for
polished electrodes of broad area.
\cite{Furuta:Nima2005,diamond1:98,Wang:2003}. Nb, Ti and Mo
electrodes were, prior to polishing, given high temperature (HT)
treatment and the Nb electrodes were treated with buffered
chemical polishing (BCP). HT treatment refers to vacuum firing in
a dedicated furnace at CERN (Conseil Europ\'een pour la Recherche
Nucl\'eaire) following CERN procedures for each material. One of
the criteria in the choice of temperature is to avoid excessive
re-crystallization \cite{Calatroni}. The temperature plateau of
the furnace is set at 900$^\circ$C for 2h for SS; Ti, Mo and Nb at
800$^\circ$C for 6h. Getter material such as Ti are placed inside
a separately pumped and degassed vacuum container to avoid any
contamination of the electrode by outgassing gas from the furnace.
In the case of Nb, after BCP, grains up to 10~mm size were
visible. HT treatment, with a sufficiently high temperature,
reduces the number of field emitters \cite{Niedermann:1986}. In
this case the temperature chosen for Nb is below this critical
temperature, hence the need for polishing. All electrodes are
measured as-received after a careful ethanol wipe before mounting
in the UHV chamber. The first PGD is made after the installed
as-received cathode either arcs or emits a dark current of 1~nA.

Table \ref{tabPolishedrslt} summarizes the best results. PGD
almost invariably improved breakdown gradient. Electrodes
installed "as-received" were only cleaned with alcohol (ethanol)
before being installed in the vacuum chamber. The values quoted
for 1~nA, were usually obtained after arcing, hence the lower
numbers compare to the $<$0.05~nA row. In the "as-received" case
the values of the field recorded in the 1~nA row are higher, as FE
appears before breakdown. After test, all electrodes retained
their mirror like appearance, except for the damaged area, however
no roughness was measured.

\begin{table}[htbp]
\centering

\caption{DC gradient in MV/m held between a cathode - anode pair
at 1~mm gap for the given dark current in nA.}
\begin{tabular}{|c|c|c|c|}
\hline Cthd-And& Dark Current  &  As Received & After PGD \\
\hline \raisebox{-1.50ex}[0cm][0cm]{Cu-Cu}& $<\  0.05$ nA& 20 & 67.5 (*) \\
\cline{2-4} & 1 nA & 25 & 67.5 (*) \\
\hline \raisebox{-1.0ex}[0cm][0.0cm]{Cu-Cu}& $<\ 0.05$ nA& 37 & 52 \\
\cline{2-4} New-Used & 1 nA &  41.9 & 47 \\
\hline \raisebox{-1.50ex}[0cm][0cm]{SS-SS}& $<\ 0.05$ nA& 65 & 71 \\
\cline{2-4} & 1 nA& 24 & 25 (0.1nA) \\
\hline \raisebox{-1.50ex}[0cm][0cm]{Nb-Nb}& $<\ 0.05$ nA& 20 & 59  \\
\cline{2-4}  & 1 nA& 23 & 20.8 \\
\hline \raisebox{-1.5ex}[0cm][0cm]{Mo-Mo}& $<\ 0.05$ nA& 32.5 (2mm) & 68.5 \\
\cline{2-4}  & 1 nA&  32 & 33 \\
\hline
\end{tabular}
\\
\begin{tabular}{c}
(*) Drift from 0.03nA to 8.5nA in 58 hrs
\end{tabular}
\label{tabPolishedrslt}
\end{table}

\begin{figure}[htb]
   \centering
   \includegraphics*[angle=-90,width=0.9\columnwidth,clip=]{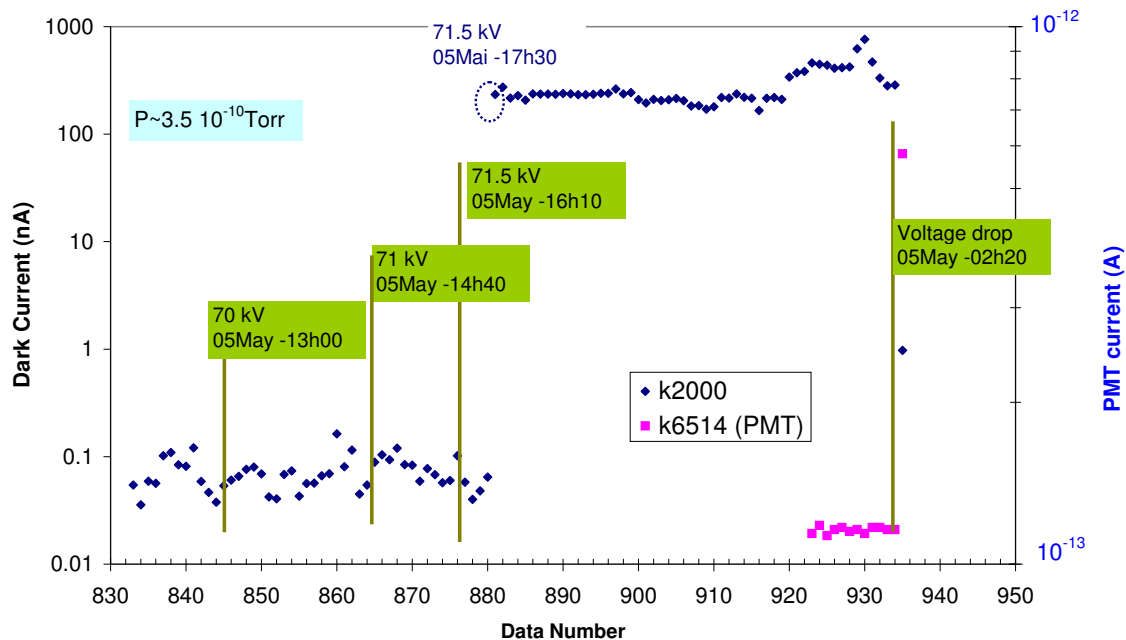}
   \caption{Dark current (label K2000) and PMT level (label K6514)
   of SS electrodes at 1~mm gap.}
   \label{figSS1mmDCdata}
\end{figure}

Operation at a given gradient was carried out over periods of
days. As an example Fig.\ref{figSS1mmDCdata} shows the results for
SS mirror-finished electrodes at 1~mm gap after PGD. Each data
point is an average of 10 measurements and was recorded at
intervals of 10~minutes. The system was stable at 69~kV for 6
days. A sudden increase of dark current was probably due to a
micro breakdown. The drop in voltage at the end of the sample is
due to a software protection scheme, which switches off the high
voltage (HV) power supply when voltage drops by more than 1~kV
from the set value. The drop in voltage is usually due to a higher
emission of dark current. The PMT signal often correlates with FE
activity, and is enhanced with increasing vacuum level. In this
setup the PMT is placed behind a window, facing the electrodes
gap. With a 4~mm gap, a clear PMT signal can be recorded, but as
the gap closes, the amount of XR going through the PMT drops
significantly.

In general, and after PGD, results obtained for the non polished
electrodes (Table.\ref{tabDCunpolished}) and polished electrodes
(Table.\ref{tabPolishedrslt}) in a DC regime are not so different.

For Nb, polishing followed by a PGD (Table.\ref{tabPolishedrslt})
permitted much higher fields to be achieved compared to non
polished electrodes (Table.\ref{tabDCunpolished}). Polished Mo did
not exceed 40~MV/m, without FE, even after two consecutive Ar PGD.
In order to reach 68.5~MV/m, the Mo cathode was run with a
6~$\mu$A, 50~kV, dark current for 72~hours. A third Ar PGD of
2~hours followed; the best result is shown in
Table.\ref{tabPolishedrslt}. The main action of the PGD is to
remove contaminants on the surface hence changing the work
function of the surface of the material with little change of the
surface topology \cite{Cernusca,springer5}

When compared to ref.\cite{Furuta:Nima2005} an in-situ PGD gives
results equivalent to those after careful polishing followed by an
ultra-pure high pressure water rinsing. In general a PGD treatment
used on a commercially mirror-finished or non mirror-finished
electrode gives significant improvement compare to an "as
Received" electrode. A literature comparison, Table
\ref{tabFuruta}, to our results proves that the breakdown field
(BF) quality from mirror-finished surfaces, either by polishing or
diamond turning, coming from commercial vendors is, nowadays, very
satisfying. Microscopy analysis does not show systematic embedded
material, roughness achievement is equivalent, and sometimes
superior to, that which a research laboratory can produce. One
surprising point is the fact that in general PGD on rough material
gives some very competitive results compare to highly polished,
hence expensive, material.

\begin{table}[htbp]
\centering \caption{Literature field gradient (MV/m) between
electrodes obtained at 1~mm gap : for 1~nA of dark current;
without FE; breakdown field (BF) with FE; CERN results; this work,
after a PGD w/wo FE. }
\begin{tabular}{|c|c|c|c|c|c|c|c|}
\hline  & SS& Cu& Ti& Mo& Mo - Ti & Al & Nb\\
\hline Furuta - 1~nA \cite{Furuta:Nima2005}& 36& 47.5& 88& 84&
103 & - & -\\
\hline Diamond - No FE \cite{diamond1:98}& - & 70 & 60 & -&
- & 85 & 92 \\
\hline Kranjec - BF \cite{Kranjec:67}& 79.8 (*) & 46.1 & 48 &
38.9 & - & 51.3 & - \\
\hline CERN - saturated BF \cite{Descoeudres:EPAC08}& 800 & 170 &
780 &
420 & - & 150 & 400 \\
\hline Tarasova - saturated BF \cite{springer5}& - & 46 & 55 &
62 & - & 45 & 48 \\
\hline Best BF (this work)& 71 & 67.5 (**) & 67 &
68.5 & - & 92 (***) & 59 \\
\hline
\end{tabular}
\\
\begin{tabular}{c|c|c}
(*) Measured at 0.5~mm & (**) 8~nA of dark current &(***) measured at 0.75~mm \\
\end{tabular}
\label{tabFuruta}
\end{table}

In early works, it was stated that the less refractory metals seem
to reach the highest breakdown field \cite{Brodie:66,Kranjec:67}.
From results presented in Table \ref{tabFuruta}, this is not so
evident. Moreover, results from CERN contradict this conclusion
with Ti, V, Cr, Mo, Nb, Ta and W holding higher gradient than soft
materials like Cu and Al, as also shown in much older work
\cite{springer5}. CERN and results cited in \cite{springer5} are
obtained by doing breakdown conditioning; at CERN between a
pointed anode and a flat cathode with gaps around 20~$\mu$m
\cite{Descoeudres:EPAC08}, and for ref.\cite{springer5} between
different kind of electrodes separated by gaps varying from 0.3~mm
to 5~mm, hence a very different modus operandi. In order to
compare data presented in line 3 of Table~\ref{tabFuruta} to other
results, the electric field in MV/m has been derived by
calculating back from the original voltage.

Different preparation will results in a different enhancement
factor ($\beta$) and a lower $\beta$ results in general in a
higher BF, although not always
\cite{Suzuki:Nima2001,Furuta:Nima2005}. Using $\beta$ is a
convenient way to compare FE quality because it is usually a
straight line fit of the data when plotted in a Fowler-Nordheim
plot (ln(I/V$^2$) vs 1/V) and should not present any discontinuity
\cite{Kranjec:67,springer5}. It is particularly well suited in
comparing electrodes of the same materials prepared in an
identical way, prior any high gradient testing. However, the
intrinsic understanding of $\beta$ on a "dirty" surface, in a UHV
sense, is absent. For example, a variation of the chemistry on the
surface implies a variation of the work function. However, as the
work function is systematically never measured, but assumed, this
variation will be attributed to a change of $\beta$ in the
Fowler-Nordheim equation. Our point is that comparing directly the
emission current versus the field applied gives an immediate
conclusion. In most of our results the FE current, before
breakdown, is below the detection limit, as in
Fig.\ref{figSS1mmDCdata}. Hence comparison is not feasible.


The electrodes tested were single material. A mix of different
electrode materials is certainly promising for reaching high
gradient in real DC gun \cite{Furuta:Nima2005}, as one of the
hypothesis of the breakdown is material transfer from the anode to
the cathode \cite{springer5}. Unfortunately our pulse has an
oscillatory behavior with a positive peak amplitude being 60\% of
the negative peak, Fig.\ref{figPulseOscillo}. As the effect of the
anode cannot be neglected, we have concentrated our effort on
single materials in order to propose usable solutions for our
needs, without complicating our research.

\section{500 kV Pulsed Diode prototype electron source}

It is known that higher gradients are achieved when using a pulse
high voltage compared to a DC system
\cite{springer5,Smedley:2003}. This is also true in an RF system
where the short pulse permits a higher gradient to be reached
\cite{Doebert:PAC07}.

\begin{figure}[htb]
    \centering
    \includegraphics*[width=140mm]{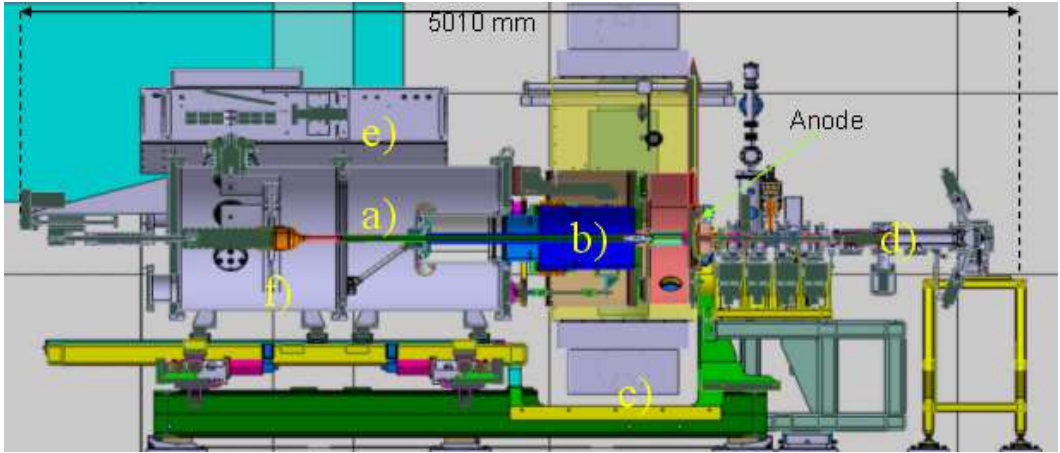}
    \caption{Cross section of the 500~kV beam line. a) SF6 Pressure tank;
     b) UHV tank, containing the 2 electrodes; c) Class 1 Clean room;
     d) Diagnostic lines containing YAG screens, Solenoids, faraday cup, and an emittance monitor
     \cite{Pedrozzi:EPAC08}; e) Thyratron; f) HV transformer }
    \label{figOBLA500}
\end{figure}

The original SwissFEL concept was based on a field emitter array
(FEA) producing an electron beam of 5.5~A in 2~ns (FWHM)
\cite{Bakker:2004}. In order to produce an electron beam with no
or little energy spread either a jitter free square pulse or a
longer pulse is needed. The 500~kV - 250~ns (FWHM) pulser,
Fig.\ref{figOBLA500}, was initially designed to accommodate such a
requirement.

The main driving factor for having HV transformer technology is
the compromise between stability and complexity. In the voltage
range considered (0.5 - 1~MV), the only available hard switch is a
gas switch (spark gap). Self-triggered gas switches are not
suitable for large dynamic range because switching parameters
depend strongly on operating voltage. Another disadvantage is
large switching time jitter and limited life. Triggered gas
switches are much more suitable in terms of jitter but require a
fast HV trigger pulse or a high power short laser pulse. Switching
characteristics of electrically triggered gas switches still
depend on applied voltage and to ensure good dynamic range and
jitter performance, they require fast trigger voltage pulses with
amplitude comparable to the switched voltage \cite{Williams:89}.
This requirement makes such a system impractical. Performance of
laser triggered gas switches depends on the properties of the
laser trigger pulse. Using lower laser peak power (several MW) the
jitter performance is still in tens of nanoseconds (peak)
\cite{Hutsel:08,Vyuga:PPC03}. Recently, research groups have
reported good results with high peak power laser triggered gas
switches \cite{Luther:2001,Hendriks}. In order to get dense enough
photo-excited plasma in the conducting channel, the trigger laser
has to provide pulses with tens of GW to TW peak power. That
dramatically increases the complexity and the cost of such a
system. Thyratron switched air-core resonant transformer, combined
with relatively high primary voltage, gives a good "speed -
stability" engineering compromise and makes it possible to achieve
the required amplitude stability to better than 0.1\%
\cite{Paraliev:IPMC08}. The linear system ensures large dynamic
range of the output voltage, required for such test system.
Another reason to use a fully linear technology is the
scalability.

\begin{figure}[htb]
   \centering
   \includegraphics*[width=0.6\columnwidth,clip=]{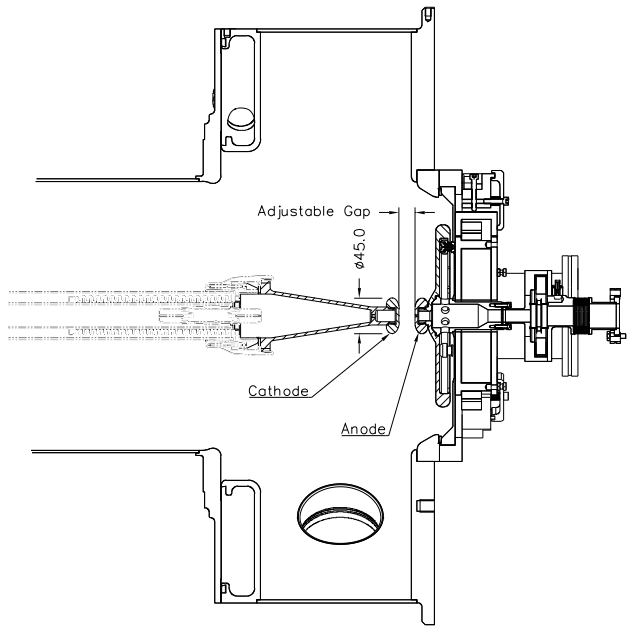}
   \caption{Diode Configuration inside the 500~kV pulser UHV chamber.
   The cathode is mounted on the left side.}
   \label{figPulserxsection}
\end{figure}

A more detailed cross section of the UHV vacuum chamber with the
electrodes for the high gradient pulsed tests is shown in
Fig.\ref{figPulserxsection}. The laser illuminating the cathode is
inserted in the middle of the beam line after the UHV valve
Fig.\ref{figOBLA500} and goes through the 2~mm diameter anode
hole. The chamber is sealed without bolts, using a differential
vacuum system. This permits easy exchange of electrodes. The
chamber is integrated into a filtered air cabinet (shown in green
in Fig.\ref{figOBLA500}) to ensure a dust-free environment when it
is opened. With a turbo pump, a pressure of 10$^{-7}$~mbar can be
reached in 12 hours, and with two 300~l/s ion pumps,
10$^{-8}$~mbar can be reached within two days. The cathode support
includes provision for future mounting of FEAs which are presently
under development at PSI \cite{Soichiro:IVNC07}. The pulser uses
thyratrons and an air-core transformer (Tesla coil) to generate a
damped oscillating waveform with a dominant negative peak voltage
\cite{Paraliev:USA05,Paraliev:IPMC08}. The peak voltage, when
above 30~kV, is stable to $\pm$0.1\% and can be varied from 0~kV
to 500~kV. The gap between the electrodes is adjustable from 0 to
30~mm. Both electrodes are made of the same material and have the
same profile, unlike the previous 100~kV DC system. For quantum
efficiency (QE) or electron beam emittance measurements, the
emitted electron beam passes through a 2~mm diameter anode hole.
The diagnostic beam line, shown in Fig.\ref{figOBLA500}, includes
five solenoids to transport and focus the beam. After the first
solenoid, which is mounted on the pulser flange, a wall current
monitor is installed. A Faraday cup or a YAG screen can be
inserted in the beam line. They are located 491~mm downstream of
the anode. An emittance monitor of 870~mm long fitted with movable
pepper pot and YAG screen is used for emittance measurement. It is
located at the end of the beam line. The overall length of the
diagnostic line from the pulser flange to the end of the emittance
monitor is 1470~mm. A more detailed description of the 5~m long
beam line, including the 3.4~m long pulser, can be found in
\cite{Pedrozzi:EPAC08}.

\begin{figure}[htb]
   \centering
   \includegraphics*[width=\columnwidth,clip=]{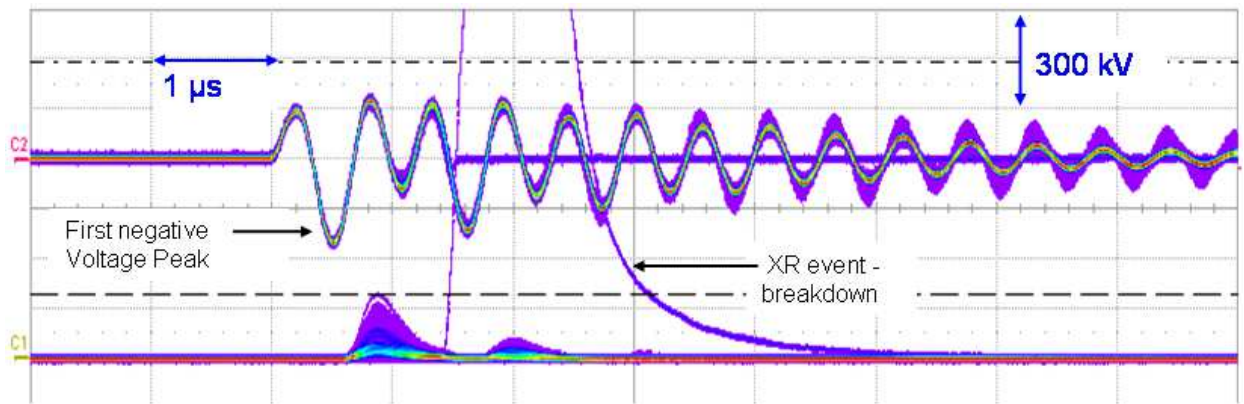}
   \caption{Measured waveforms of the HV pulse -280~kV (first negative peak) followed by a +170~kV
   (second positive peak)(upper trace, C2) and X-ray scintillator signals lower trace, C1.}
   \label{figPulseOscillo}
\end{figure}

Fig.\ref{figPulseOscillo} shows a typical pulser trace with a 4~mm
gap. The upper trace labeled C2 is a collection of waveforms of
the quasi DC pulse. The pulse oscillates up to 10~$\mu$s; 8~$\mu$s
are displayed in Fig.\ref{figPulseOscillo}. This oscillation, and
the superposition of a 950~kHz and a 1800~kHz signal, results from
the choice of technology \cite{Paraliev:USA05}. The amplitude of
the second positive peak is 60\% of the maximum of the first
negative peak.

\begin{figure}[htb]
   \centering
   \includegraphics*[width=0.7\columnwidth,clip=]{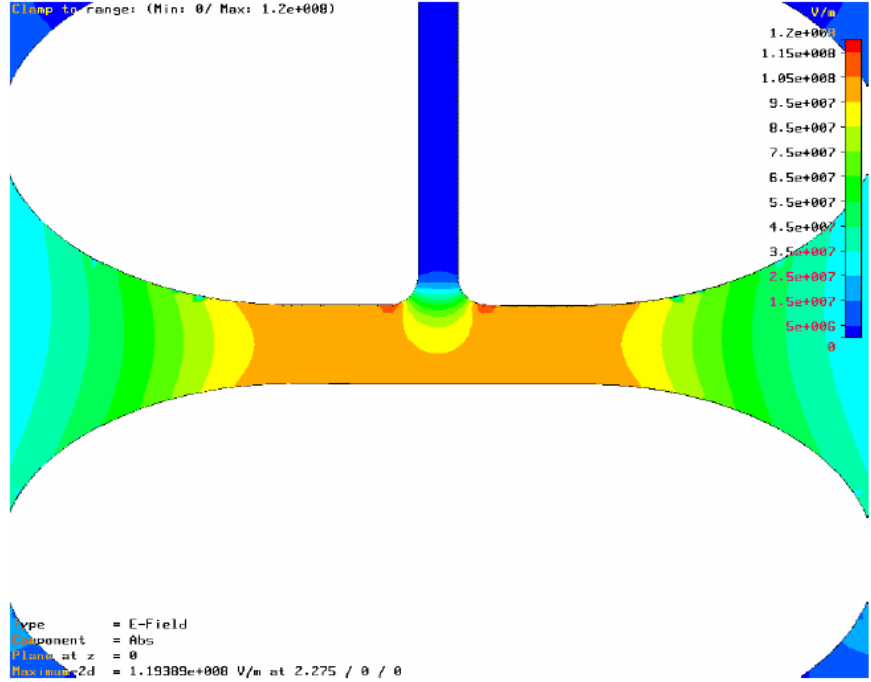}
   \caption{Electric Field simulation of pulser's electrodes showing a peak value at the anode surface of 119~MV/m for
   4~mm gap and 400~kV applied voltage}
   \label{figEFsimulation}
\end{figure}

The oscillatory voltage waveform of the pulser, with still a high
amplitude of the positive peak, was the reason to concentrate our
effort on electrode pairs of the same material in the 100~kV DC
test stand. Fig.\ref{figEFsimulation} shows the electric field at
different locations of the pulser electrodes' surfaces.

An XR scintillator detector is used to detect breakdown. The XR
scintillator signals, multiple traces, labeled C1 are delayed with
respect to the negative voltage peaks because of an analog filter
delay. When a spark occurs, the XR signal saturates as is seen for
one event in Fig.\ref{figPulseOscillo}; this high level is used as
a machine protection system and will stop the HV pulsing until
reset, or at least until the HV is reduced. Breakdown is
characterized by watching the camera pointing toward the
electrodes, the cathode voltage and XR signals. Considering
Fig.\ref{figPulseOscillo} in more detail, the spark occurred at a
moment when the HV was low, suggesting that the previous positive
voltage swing may have contributed. The HV is clamped to a flat
line with almost zero voltage indicating that all energy is
dissipated in the spark turning it into a destructive arc. The
video camera image shows a bright uniform arc column between anode
and cathode and the scintillator signal saturates for a few
$\mu$s. In general, more than 2/3 of the events happen at the
first negative voltage peak. The scintillator will saturate, but
the pulser will not systematically display a flat line.

A total of 24 electrode pairs were tested, with evolving skill for
polishing and cleaning. The reason for developing in-house
polishing technics, instead of buying pairs from vendors, is from
the possibility of high turn over in electrode exchange. The range
of breakdown gradients achieved, for gaps varying from 2~mm to
6~mm, is shown in Table \ref{tabpulsedresult}. Highest gradient
were obtained for the smallest gap. For reasons of cost and ease,
most measurements were obtained with 316L stainless steel and OFE
copper, with typical average roughness below 15~nm. Although many
factors might affect breakdown, the dominant factor for this
gradient level is the quality of mechanical polishing, mainly
characterized by roughness and some microscopy to pinpoint
embedded polishing material. The hand polished procedure starts
with the use of abrasive paper from coarse to fine grains. For SS,
diamond paste of decreasing diamond particles size is then used.
For Cu, polishing paste LUXOR$^{TM}$ 0.1Mu is used, instead of
diamond paste; wipe of the paste is performed using ethanol.

Before closing the vacuum chamber all electrodes are spray cleaned
with dry ice, for a few seconds. First results on Nb were obtained
with an electrode prepared at CERN, using the BCP technique, which
again exposed the big grains of the material. Higher results
(83~MV/m) were obtained after hand polishing, which did not expose
the underlying grain structure of the electrode so completely
hence reducing the roughness. SS and TiVAl achieved the highest
gradient although they are alloys. It seems that material purity
is not an important parameter. So far, we have not seen any
influence of the electrode geometry, e.g. the anode hole, on
breakdown rate or breakdown location.

\begin{table}[htp]
\centering \caption{Summary of peak gradient (MV/m) for gaps
ranging between 2~mm and 6~mm.}
\begin{tabular}{|c|c|c|c|}
\hline Cthd-And & without laser  & Duettino$^\circledR$ laser & Jaguar$^\circledR$ laser\\
\hline {CuCr - CuCr}&  23 & - & -  \\
\hline {WCr - WCr}& 30  & - & - \\
\hline  {Al - Al}& 46 & - & 50 \\
\hline  {Cu - Cu}& 74 - 83 & 41 - 51 & 50 - 63\\
\hline  {Nb - Nb}& 60 - 83 & 64 & -\\
\hline  {SS  - SS}& 70 - 140 & 95 & 50 - 80 \\
\hline {TiVAl - TiVAl}& 70 - 130 & - & -\\
\hline
\end{tabular}
\label{tabpulsedresult}
\end{table}

\begin{figure}[htbp]
\begin{minipage}[t]{.45\linewidth}
\centering
\includegraphics[clip=,width=\columnwidth,totalheight=6cm]{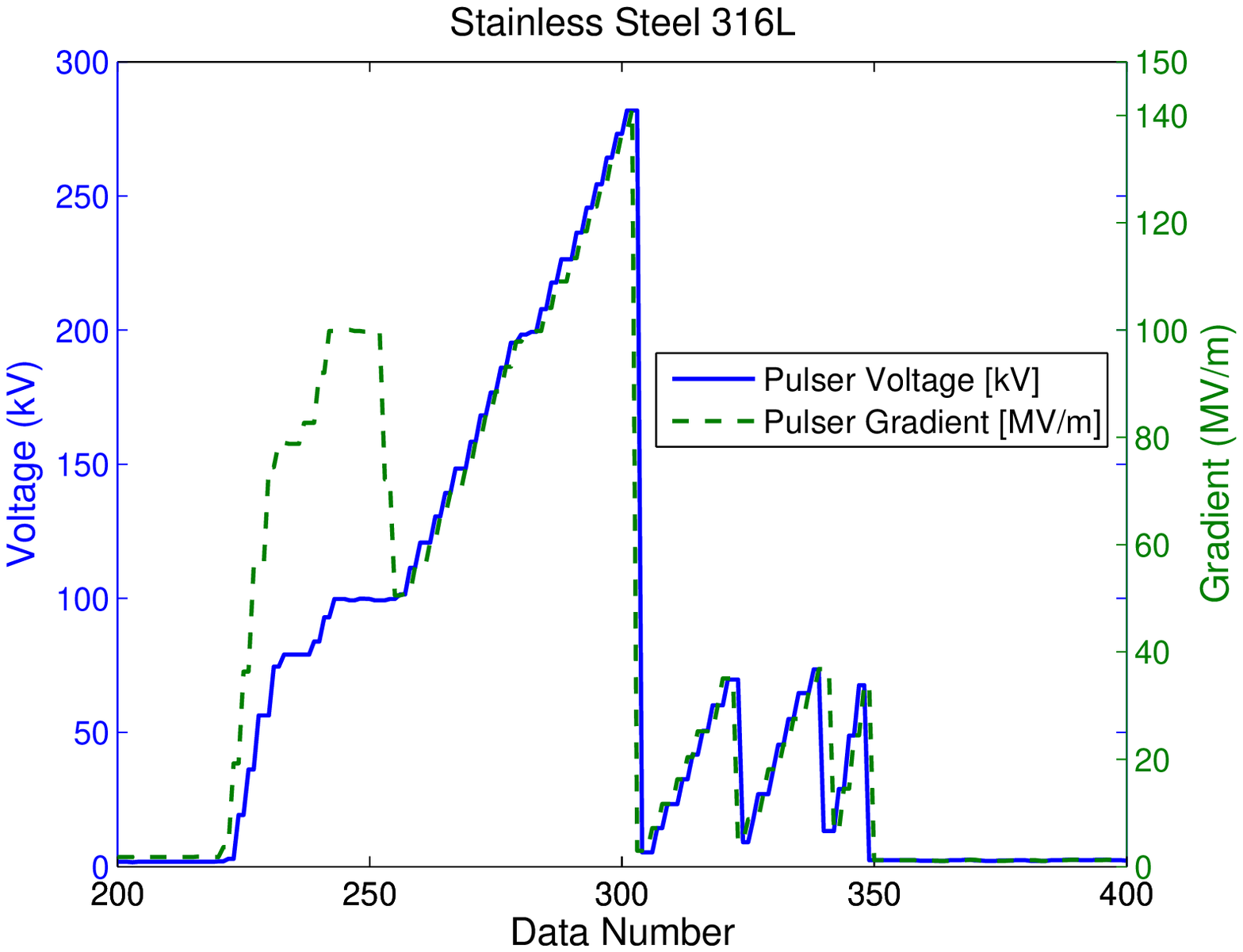}
\end{minipage}
\begin{minipage}[t]{.45\linewidth}
\centering
\includegraphics[clip=,width=\columnwidth,totalheight=6cm]{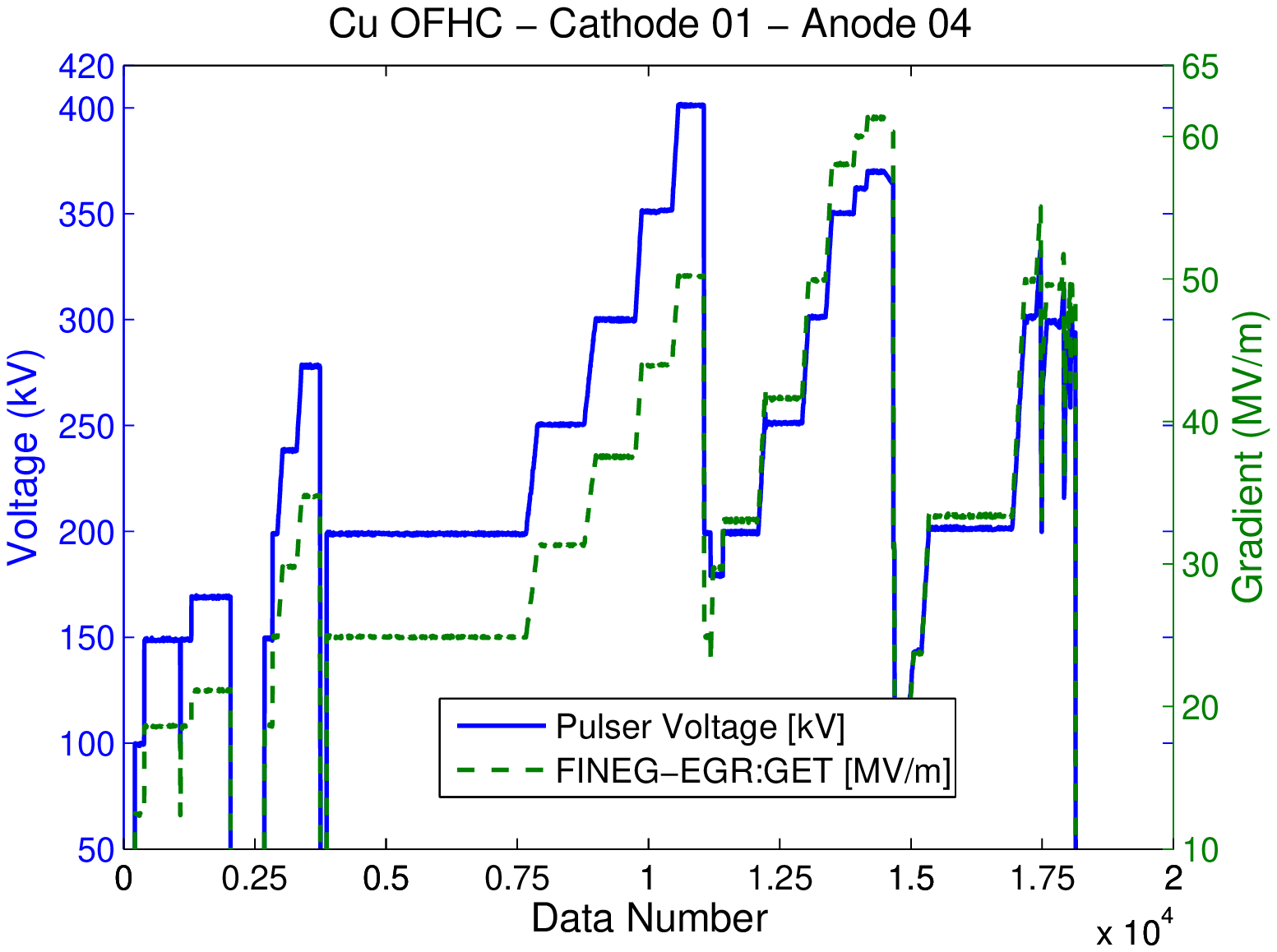}
\end{minipage}
\caption{Operation with SS 316L and Copper OFE electrode. Left
Figure : Typical operation for HG without laser. Right Figure :
Typical operation with laser.} \label{figHVHGoperation}
\end{figure}

Typical operation for HG achievement, without laser, is shown in
Fig.\ref{figHVHGoperation} (left figure). The voltage (solid line)
for a given gap is raised in steps of a few kV after a given
number of pulses. The voltage increases until a breakdown occurs.
At this moment the voltage is decreased and ramping up is resumed.
This procedure is used with and without laser operation. In this
example, the hand polished SS electrodes did not survive the first
breakdown, as gradient could not be restored by subsequent
breakdowns. Operation lasted for 31~minutes. The right figure of
Fig.\ref{figHVHGoperation} displays a typical laser operation.
First at large gap then at smaller gaps. This procedure is the
standard procedure used for all electrodes illuminated with a
laser. The corollary of this procedure is that an electrode
holding high voltage at small gap will hold the same voltage at
large gap. We routinely observe this behaviour each time we resume
operation, day to day, with the installed electrode pair. The
constant voltage plateaus signify gap change or QE measurement. In
this example, the diamond turned electrodes broke down at 61~MV/m
and the gradient could not exceed 50~MV/m. Operation lasted for
10~hours. The longest operation at high gradient, without
breakdown, was obtained with SS electrodes : 80~MV/m for 9 hours
and extracting 55~pC of electron beam current. Routine operation
for emittance measurement at 50~MV/m (6~mm gap) over days, is
achieved with either Cu or SS electrodes. The simple hand
polishing technique used can allow a damaged electrodes to be
refurbished in half a day and be able to be processed (in general
discharge free) to 50~MV/m.

\begin{figure}[htb]
   \centering
   \includegraphics*[angle=0, width=0.95\columnwidth,clip=]{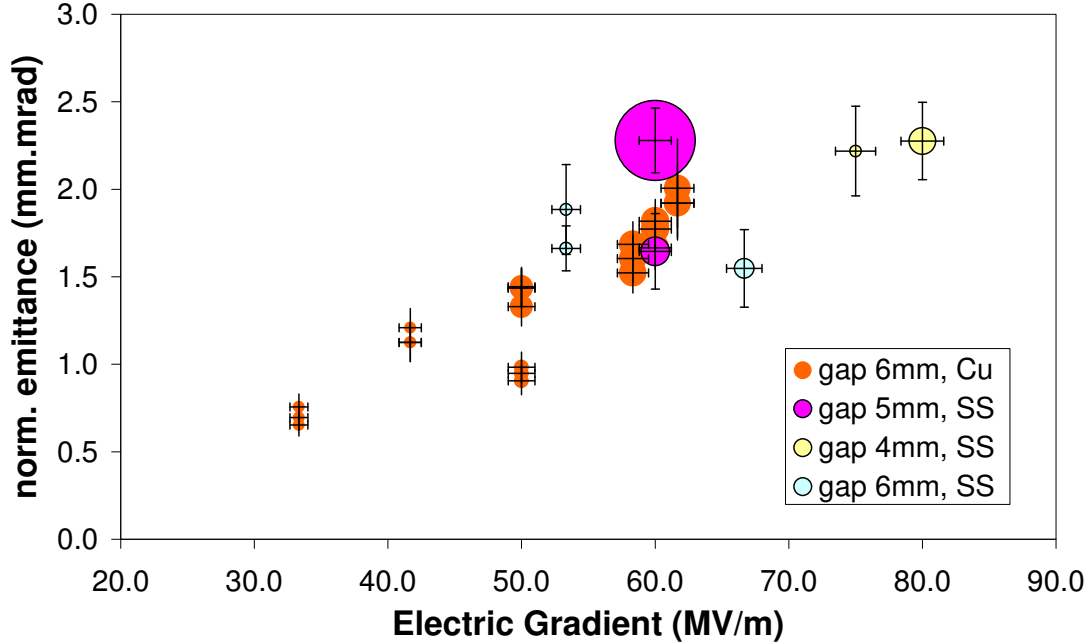}
   \caption{Normalized transverse emittance in function of the charge (colored circled)
   and the electric gradient for Cu and SS electrodes \cite{Oppelt}}
   \label{figNormEmittc}
\end{figure}

The charge produced by the laser lies between 1~pC to 80~pC
depending of the needs. The goal of this electron gun, in its
original concept, is to produce a low emittance beam at high
voltage and gradient. During cathode testing, no systematic
emittance studies were done, but still some normalized transverse
emittance (geometric average $\sqrt{x.y}$) as function of the beam
energy were measured, Fig.\ref{figNormEmittc} \cite{Oppelt}. The
beam charge, represented by the circle size, ranges from 10~pC to
154~pC and most emittances measured lie between 0.5~mm.mrad and
2~mm.mrad. Emittance measured with Cu cathodes shows that in
general smaller emittance were obtained for small charges. We
could not correlate small emittance for high gradient at similar
charges, orange circles in Fig.\ref{figNormEmittc}. Results on SS
were even more inconclusive.

In the 100~kV DC system PGD have shown to be pushing back the
onset of the breakdown threshold, even on mirror surfaces. In the
pulser we used an external DC power supply to bias the cathode
positive or negative in order to light up ArPGD. The plasma could
be localized between the electrodes only for a given set of
voltage (200~V), current and gas pressure (usually a few Torr),
and this was largely independent of the gap between the
electrodes. With such parameters and configuration, the energy of
the ions was probably much less than 200~eV. Later attempts used
the pulser itself, running at 50~Hz to give a kilovolt pulse
across the gap. Unfortunately, the large gap current meant that
the cathode had only a positive half-cycle of some microseconds.
PGD treated electrodes did not perform better than untreated
electrodes.

As expected, the gradients and voltages achieved for the pulsed
diode system (Table.\ref{tabpulsedresult}) with gaps varying from
2~mm to 6~mm are above the one obtained in DC system with large
gap (1~mm to 4~mm) (Table.\ref{tabFuruta}). Moreover DC breakdown
test using pulser's electrodes design, Fig.\ref{figEFsimulation},
with SS and Cu electrodes, prepared similarly as the one installed
in the pulser, holding the dry ice spray, have reach gradient
similar to gradient quoted in Table.\ref{tabFuruta}, respectively
69~MV/m and 63~MV/m at 1~mm gap.


\section{Quantum Efficiency of Metals}

A collection of quantum efficiency, or photoelectric yield curves
for "dirty surfaces", in the UHV sense, and for photon energy
above 10~eV, is summarized in \cite{Redhead:68}. QE for clean and
"dirty" copper for wavelength above 20~nm can be found in
\cite{Phelps:1999}.

Two laser systems, from Time-Bandwidth \cite{Timebandwidth}, were
used to measure the QE of the metals. The first laser system used
to measure the QE of the material is described in
\cite{Pedrozzi:EPAC08}. The main parameters of the
Duettino$^\circledR$ laser beam are : $\sigma$ = 6.5~ps rms,
$\lambda$ = 266~nm (4.66~eV), energy 12~$\mu$J/pulse at laser
exit; energy at the entrance of the accelerator beamline 4~$\mu$J
($\pm$~5\% peak to peak). The main parameters of the second laser
system, Jaguar$^\circledR$, are : $\sigma$  = 13~ps rms (or 35~ps
FWHM), $\lambda$ = 262~nm (4.73~eV), energy 400~$\mu$J/pulse at
laser exit; energy at the entrance of the accelerator beamline
200~$\mu$J ($\pm$~0.5\% peak to peak) without any laser beam
shaping pinhole. The short term laser pointing stability, for both
lasers, at a virtual cathode position was measured to be around
10~$\mu$m rms. The size of the laser beam on the cathode is chosen
such that its power density stays below 150~MW/cm$^2$, typically
the beam size is in the order of a mm when using the
Jaguar$^\circledR$ and was halved when using the
Duettino$^\circledR$.

\begin{figure}[htb]
\vspace{-0.5cm} \setcaptionwidth{6.5cm}
\begin{minipage}[t]{.45\linewidth}
   \centering
   \includegraphics*[width=\columnwidth,totalheight=7cm, clip=]{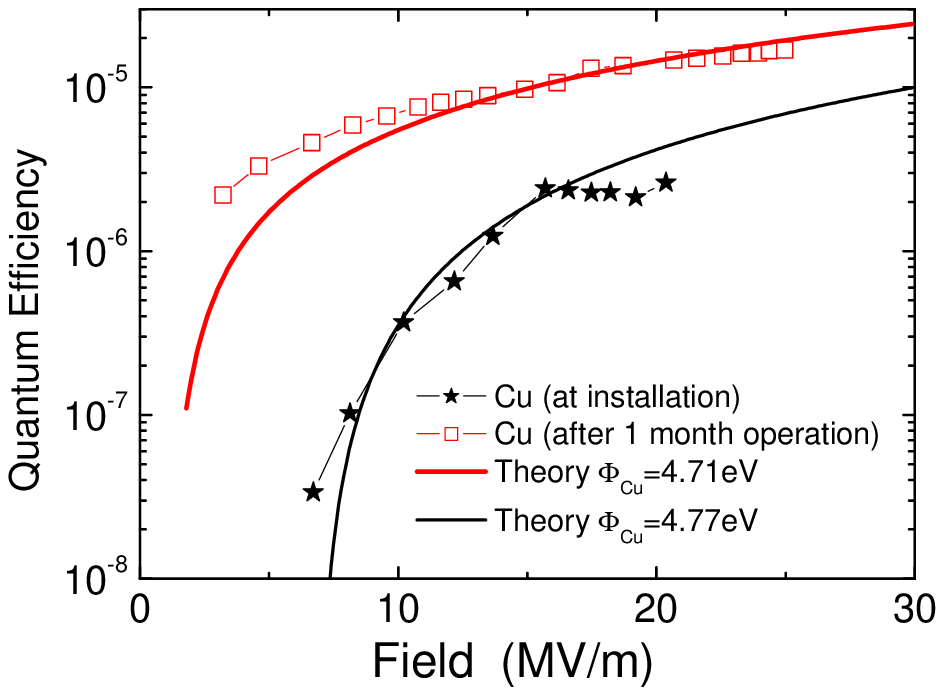}
   \caption{QE increase for a Cu cathode after 1 month of laser and high voltage operation.}
   \label{figQE02}
\end{minipage}
\begin{minipage}[t]{.45\linewidth}
   \centering
   \includegraphics*[width=\columnwidth,totalheight=7cm, clip=]{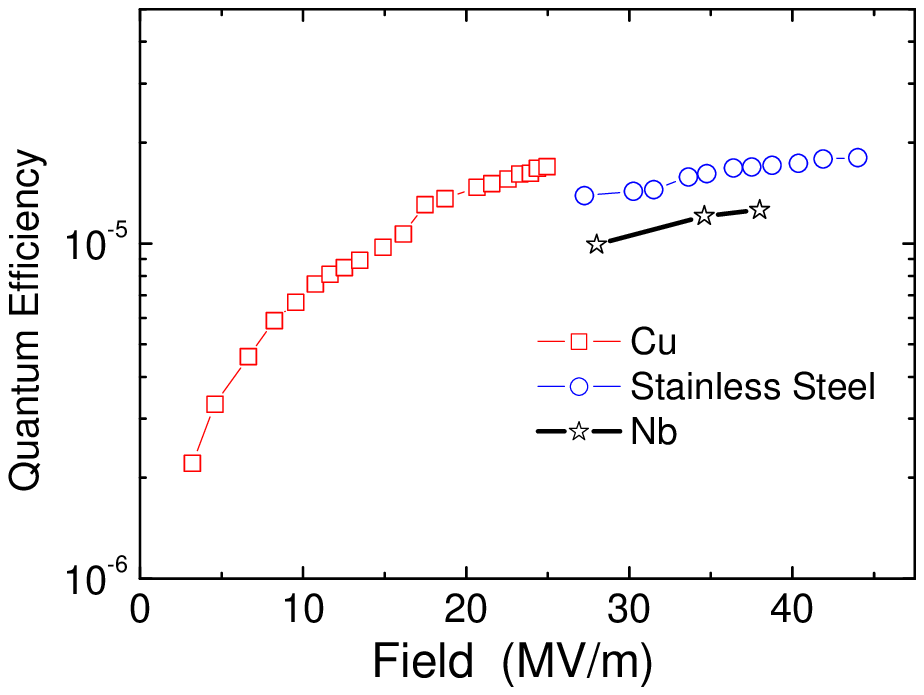}
   \caption{QE of copper, stainless steel and niobium}
   \label{figQE01}
\end{minipage}
\end{figure}

As represented in Fig.\ref{figQE02}, the quantum efficiency (QE)
increases with the applied electric field due to the Schottky
effect.

Indeed analytical formulae \cite{Dowell:2006} based on the three
steps model (absorption, excitation and emission of electrons)
\cite{Dubridge} reproduce quite well the measured data points. The
only parameter in equation~\ref{EquQE} that has been adjusted is
the work function, $\Phi$ which depends mainly on the surface
material and crystal orientation. The reflectivity R, of a given
metal, is measured in air.

\begin{eqnarray}
QE(\omega) = \frac{1 - R(\omega)}{1 + \frac{\lambda_{opt}}{2 \
\lambda_{e-e}(E_m)} \ \frac{E_{ph}\sqrt{\Phi_{eff}}}{E_m^{3/2}}
(1+\sqrt{\frac{\Phi_{eff}}{E_{ph}}})} \times \nonumber \\
\frac{E_F +E_{ph}}{2E_{ph}} \times \left[1+ \frac{E_F +
\Phi_{eff}}{E_F + E_{ph}} - 2\sqrt{\frac{E_F + \Phi_{eff}}{E_F +
E_{ph}}}\ \right]
 \label{EquQE}
\end{eqnarray}

Where E$_F$ is the Fermi energy, E$_{ph}$ = $\hbar\omega$ is the
photon energy, $\lambda_{opt}$ is the laser penetration depth,
$\lambda_{e-e}$ is the electron - electron scattering length and
E$_m$ is the energy above the Fermi level. $\Phi_{eff}$ is the
effective work function, which is the work function $\Phi_{0}$ of
the bare material minus the barrier reduction due to the external
field applied, see equation \ref{EquWorkFunction}

\begin{eqnarray}
\Phi_{eff} \ (eV) = \Phi_{0} - \sqrt{\frac{e \ E (V/m)}{4 \pi
\epsilon_0}} \label{EquWorkFunction}
\end{eqnarray}

The plot of equation~\ref{EquQE} in Fig.\ref{figQE02} shows a good
match with the experimental values. The parameters used for copper
\cite{Dowell:2006} are : R$_{Cu}$(266~nm) = 0.25; E$_F$ = 7~eV;
E$_{ph}$ = 4.6~eV; $\lambda_{opt}$ = 10.7~nm; $\lambda_{e-e}$ =
2.2~nm.

The difference between the two Cu measurements can be interpreted
as a cleaning (contaminants removal) of the surface after one
month of operation; no breakdown occurred during that period. The
maximum QE measured was on the order of 2.10$^{-5}$ at 25~MV/m,
which is in agreement with copper based RF photogun measurements
\cite{Akre:2008}. Fig.\ref{figQE01} shows the QE of various metals
measured with this beamline \cite{Pedrozzi:EPAC08}.

Laser cleaning can also be efficient if the power density stays
below the ablation threshold, in order to preserve emission
uniformity at the surface. Some experimental results can be found
in \cite{Akre:2008,kirby:QE06}. As in the DC system, where ion
cleaning (PGD) has improved the field holding, ion cleaning can
also improve the QE \cite{Dowell:2006}. The effect being mostly to
reduce the work function, by removal of the oxide of the surface,
to equal the work function of the pure material, without
drastically changing the surface morphology. Arcing changes the
chemistry and topology of the surface; its effect on the QE is
seen as soon as the pulser system recovers from the arc. The
tendency is that the QE is increased after an arc. Breakdown
conditioning will definitively clean the surface and is an
efficient process for RF accelerator structure conditioning, but
is allegedly detrimental for an electrode which should produce a
uniform beam.

\begin{figure}[htbp]
\begin{minipage}[t]{.45\linewidth}
\centering
\includegraphics[clip=,width=\columnwidth,totalheight=6cm]{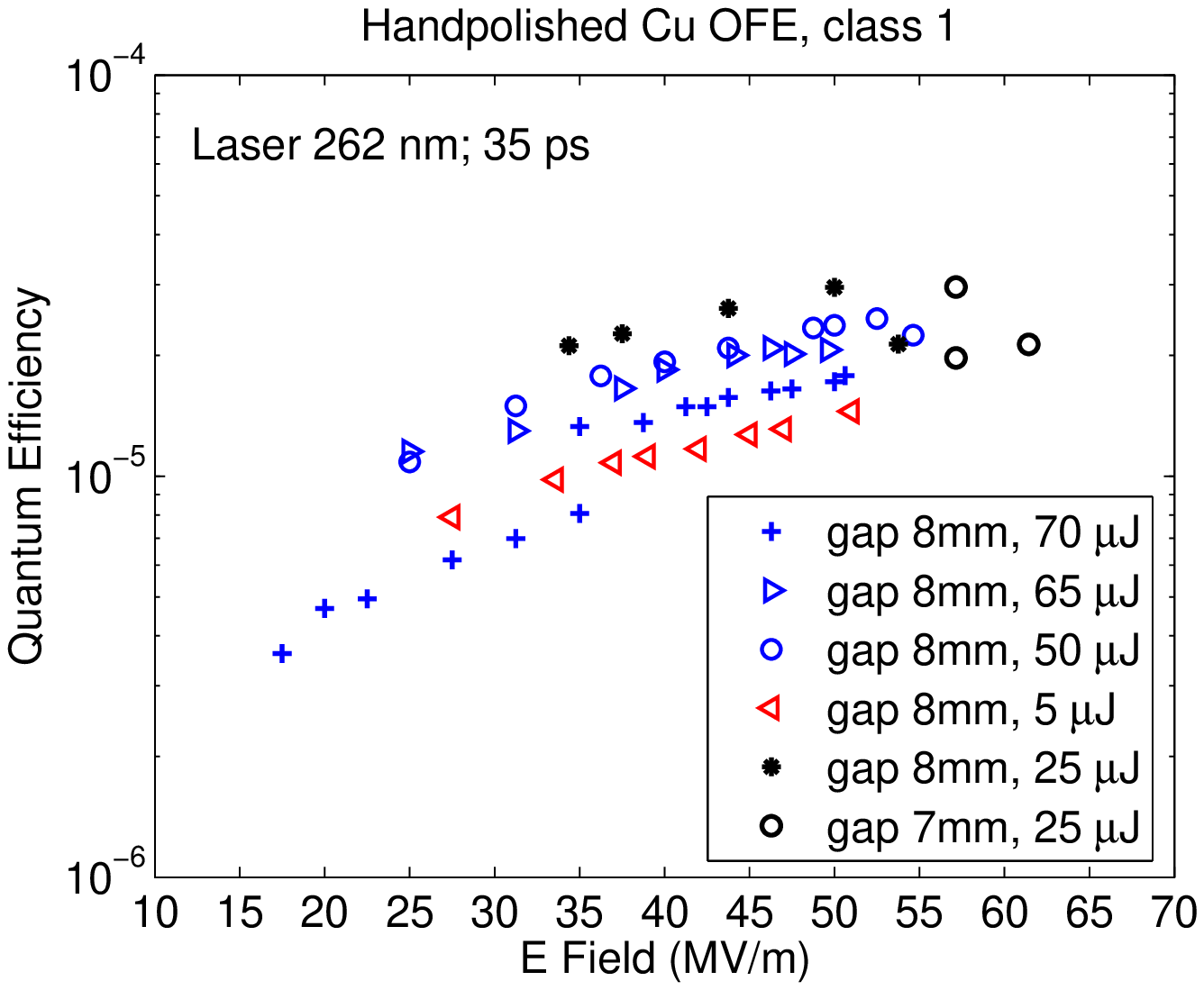}
\end{minipage}
\begin{minipage}[t]{.45\linewidth}
\centering
\includegraphics[clip=,width=\columnwidth,totalheight=6cm]{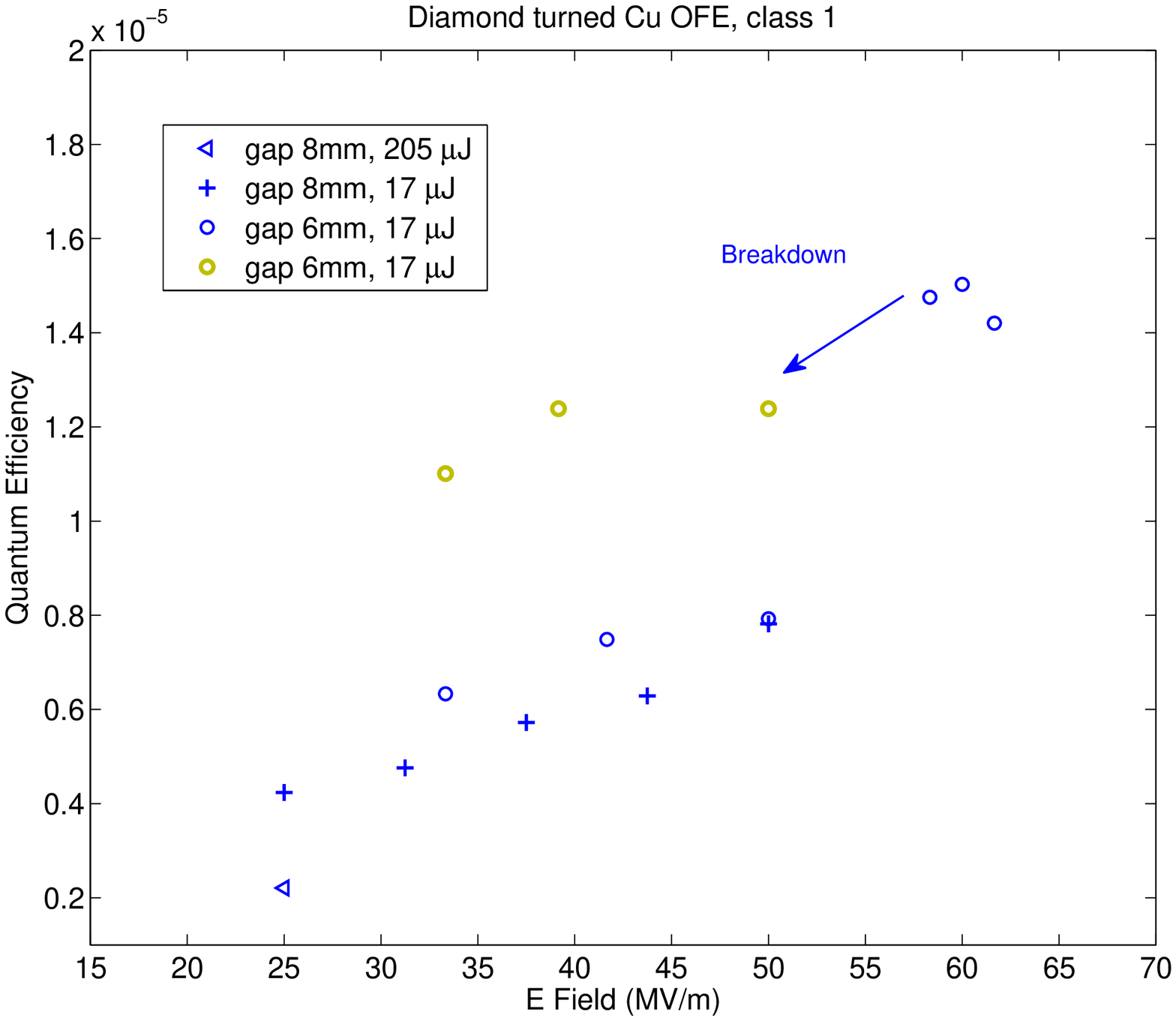}
\end{minipage}
\caption{QE of Cu OFE class 1 copper after pre-high gradient test
(83~MV/m), left figure, and a pristine pair of Cu electrode, right
figure} \label{figQECu}
\end{figure}

\begin{figure}[htbp]
\begin{minipage}[t]{.45\linewidth}
\centering
\includegraphics[clip=,width=\columnwidth,totalheight=6cm]{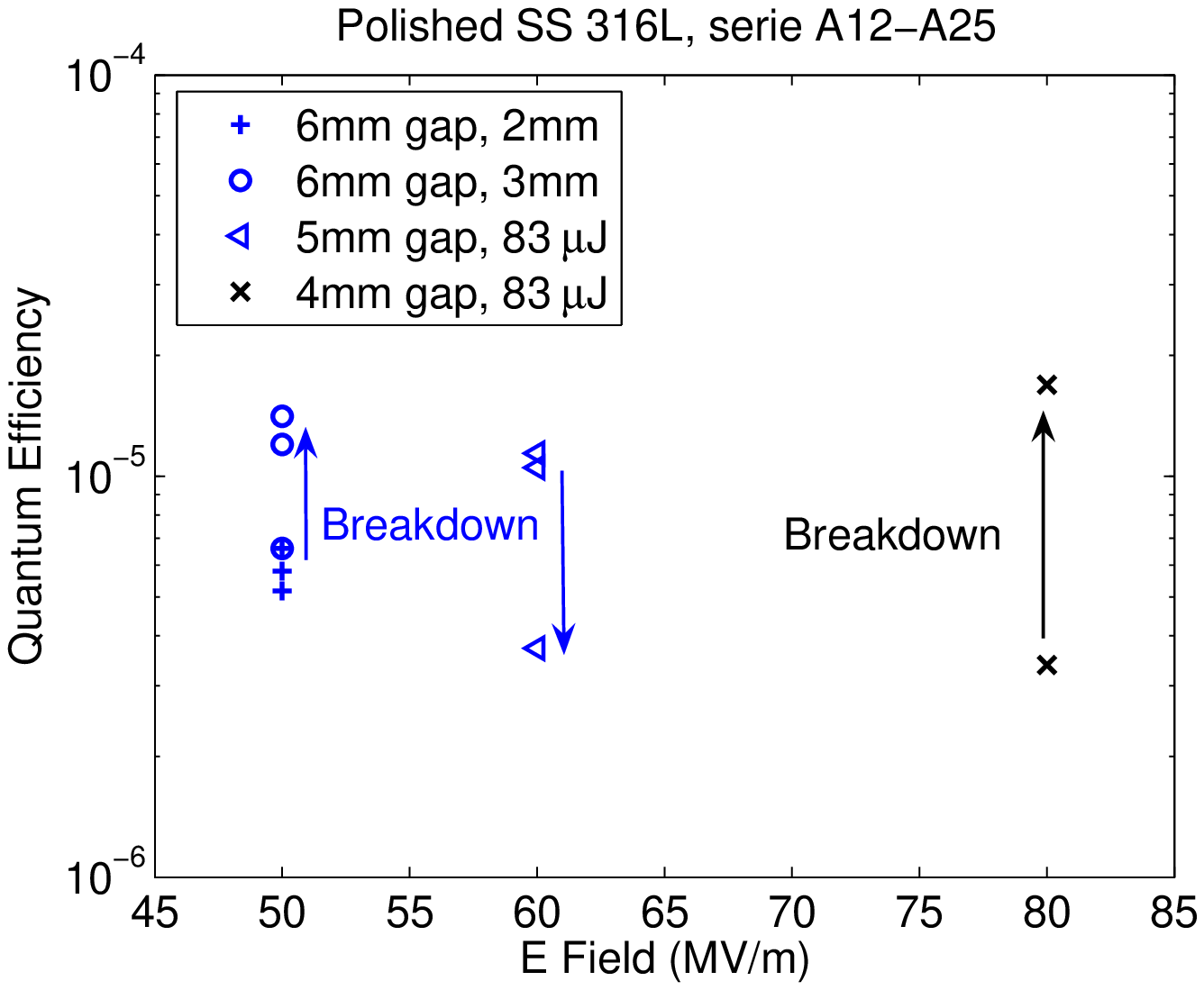}
\end{minipage}
\begin{minipage}[t]{.45\linewidth}
\centering
\includegraphics[clip=,width=\columnwidth,totalheight=6cm]{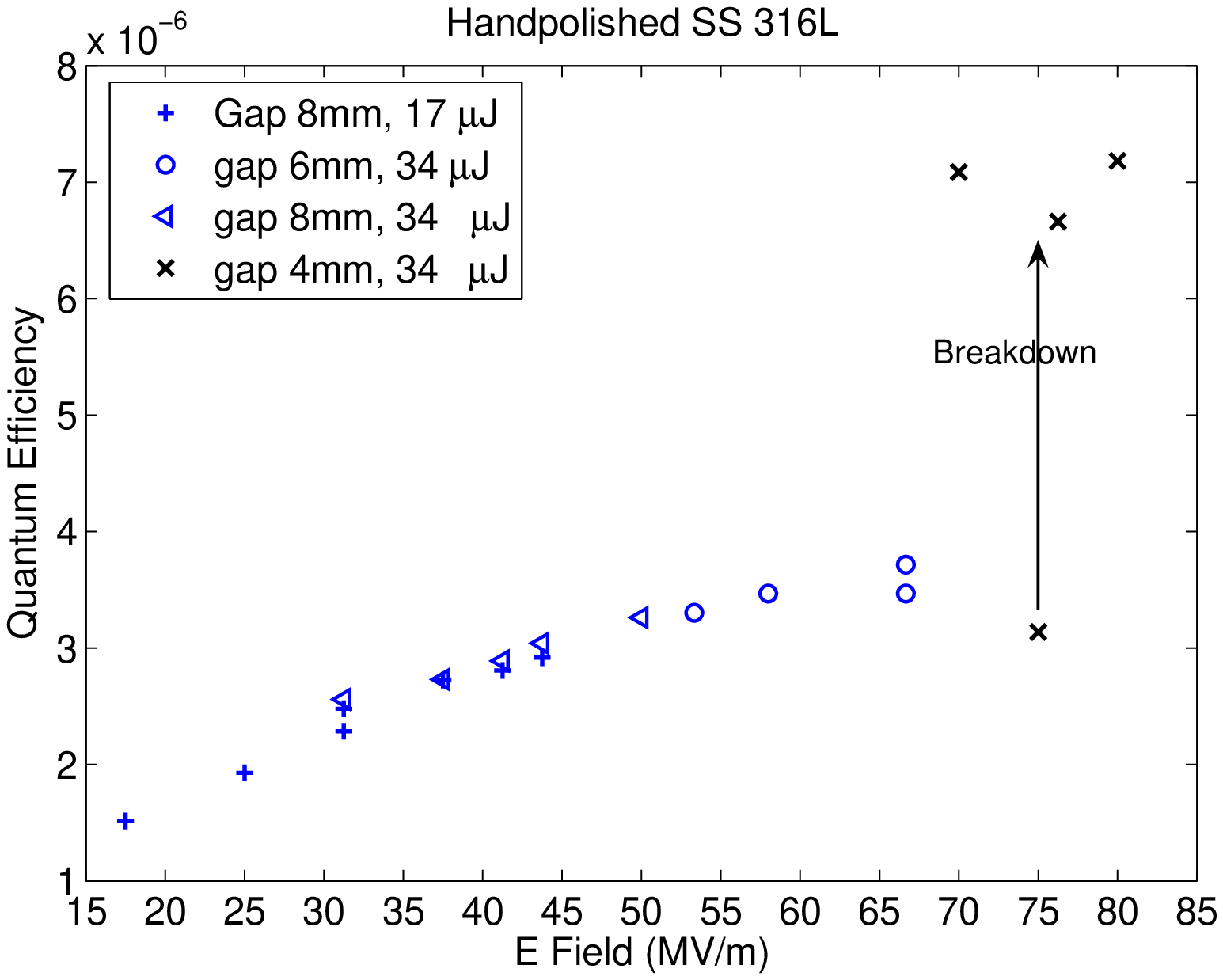}
\end{minipage}
\caption{QE of pristine SS electrodes, from a commercially
polished, left figure and hand polished, right figure.}
\label{figQESS}
\end{figure}

The following figures illustrate the effect of breakdown on the QE
of Cu, Fig.\ref{figQECu}, and SS, Fig.\ref{figQESS}. The data were
obtained using the Jaguar$^\circledR$ laser.

\begin{figure}[htb]
   \centering
   \includegraphics*[width=0.55\columnwidth,clip=]{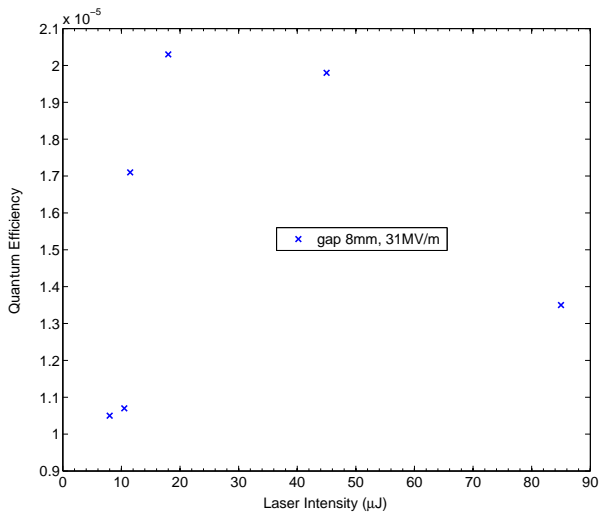}
   \caption{QE vs Laser intensity, at 262~nm, of a diamond turned Cu for a constant gradient}
   \label{figQEvsLaser}
\end{figure}

The QE displayed in the left figure of Fig.\ref{figQECu}, were
obtained using a pre high gradient (83~MV/m) conditioned pair of
Cu electrodes. On the right figure, pristine diamond turned Cu
electrodes were installed. 200~pC could be extracted using
50~$\mu$J of laser intensity. Our electron beam with charges of
10~pC and above is space charge dominated. The collection of all
charges on the Faraday cup might not be linear with the laser
intensity for gradients below 40~MV/m. Depending of the laser
intensity falling on the cathode the electron emission can be
hampered by the field produced by the already emitted electron,
even using ps long laser pulse. In Fig.\ref{figQECu} (right plot),
the QE of a pristine Cu cathode, at 25~MV/m electric field, is
twice less when using 205~$\mu$J of laser energy than when using
17~$\mu$J. We tested the QE of the same diamond turned Cu cathode
versus the laser energy, Fig.\ref{figQEvsLaser}, at 31~MV/m after
the electrode pair had sustained few breakdowns. Again for laser
intensity in excess of 40~$\mu$J, the process of electron
production became inefficient. For gradient in excess of 60~MV/m
linearity with the laser intensity up to 80~$\mu$J is ensured. For
laser intensity up to 34~$\mu$J, QE linearity is insured from
field above 35~MV/m, as can be seen in Fig.\ref{figQESS}, right
plot, as the QE is identical for two laser intensities 17~$\mu$J
and 34~$\mu$J. The QE of pristine SS electrode is slightly
inferior, for a given gradient, to the QE of a pristine Cu; see
Fig.\ref{figQESS} right figure compared to Fig.\ref{figQECu} right
figure. As for Cu, breakdown generally increases the QE and spoils
the emittance of the electron beam. However, as for Cu, a few
breakdowns ($<$~5) are usually necessary to significantly degrade
the beam emittance.

Finally the 262~nm UV light reflectivity of some electrodes were
measured, in air, after breakdown test. The light was collected at
an angle of 30~degrees. The reflected light in the damaged area
(craters) was still very specular except for Nb where strong stray
light could be observed. One should then be careful in estimating
the QE of a damaged photocathode with equation.\ref{EquQE} as we
did not collect the light over all angles.

\begin{table}[hbtp]
\centering \caption{Reflectivity of metal photocathode at 262~nm
wavelength}
\begin{tabular}{|c|c|c|c|c|}
\hline & Copper  &  316L SS & Molybdenum & Niobium \\
\hline Pristine area&  25.4\% & 17\% & 40.5\% & 12\% \\
\hline Damaged Area & 6.4\%  & 7.3\% & - & 10.5\% \\
\hline
\end{tabular}
\label{tabReflectivity}
\end{table}

\section{Conclusion}

In the quest to find a good material, capable of holding high
gradient ($>$ 100~MV/m) and producing a 200~pC electron beam; we
have tested electrodes of different material and with different
preparation procedures. We mainly compared polished and diamond
turned (for non ferrous material) electrodes to non polished one
with and without plasma glow discharge. Those results were also
tested versus results found in the literature. The main criteria
being that the material should reach the target gradient and
voltage without arcing. Any arcing being detrimental to the
emittance of the beam to be produced or could transfer foreign
material onto the electron source, effectively deteriorating this
one. This is especially true for metals of low melting point like
Aluminium.

The important results of the first section is that polishing
($<$40~nm average roughness) does not give one much advantage in
the attainment of high gradient compared to rough electrodes
($<$200~nm average roughness) which have been cleaned in-situ by
an Ar PGD. However, in the absence of PGD, polishing and good
cleaning before mounting is mandatory. Nowadays industrial
companies do produce electrodes of equivalent surface finish
(geometrically and chemically) than in-house production, with
suitable and reproducible results for our high gradient needs. The
choice to either develop in-house a polishing recipe or buy
polished electrodes from external vendor, depends on the turn over
time and price to obtain new electrodes and not on the quality of
gradient achieved as they both gave similar end results. The PSI
in-house recipe, which mainly depended on human skill, took two
month to master and allowed us to refurbished an electrode pair in
half a day.

The overall high gradient results obtained can be directly of use
for DC guns like the 500~kV DC gun used at Jefferson Laboratory, a
gun design which is now widespread in the FEL/ERL community
\cite{Sinclair:2006}. In this remark we are of course not
suggesting that a steel cathode holding 69~MV/m at 1~mm under DC
voltage will hold the same gradient at 10~mm. We are instead
suggesting material and possible preparation methods to be used
for those guns. One has still to remember that using the "same"
preparation technics does not, unfortunately, necessarily implies
the same end results. However, this guarantees still good
performances, on routine operation.

In the second part of this work, higher gradients have been
obtained using a voltage pulse of 250~ns (FWHM). We have reached
the highest gradient with a Ti alloy and SS.  Sustained operation
with hand polished SS electrodes at 80~MV/m with 55~pC of electron
beam extracted was proven for 9~hours. It is not understood why
the electrode eventually arced, as it is often the case in this
field. Operation, using electron beam with various charges, at
$\sim$300~kV, $\sim$50~MV/m for days without breakdown and no
measurable dark current is now routinely achieved, with either SS
or Cu, polished, electrodes. In most cases, those electrodes did
not arc during the voltage increase. Dark current do exist, as
shown by the XR scintillator activity, but the beam is not
captured and transported along the beam line.

Cu and SS were found to have QE in the 10$^{-6}$ range when
pristine and for field below 80~MV/m. Breakdown usually rises the
QE a decade higher, sometime 2 decades, but degrades the quality
of the electron beam. The QE stays constant using different laser
energy hence the emitted current scales linearly with the laser
energy. This will be true as long as the electric field applied
balances the space charge of the extracted beam.

UV (262~nm or 266~nm) reflectivity of the electrodes drops on
damaged areas compared to pristine ones. Rough areas trap more
light, which makes the surface look black under visible light,
hence increasing the QE. Finally, laser ablation threshold from
metals depends strongly on the pulse length. For UV laser pulse of
$\sim$30~ns, the fluence threshold is $\sim$J/cm$^2$
\cite{Ashfold:2003,Lorusso:2008}. For picosecond laser pulse the
threshold starts around a few 100~mJ/cm$^2$
\cite{Preuss:1995,Schafer:2002}. Our first attempts to focus the
Jaguar$^\circledR$ laser beam on a polished Mo cathode, in vacuum
and in presence of high voltage, using a beam radius of 300~$\mu$m
and 75~$\mu$J of incoming UV light, resulted in a hole 100~$\mu$m
deep in the metal. Gradient did not exceed 18~MV/m (100~kV). It
seems that to keep the electrode safe from ablation and keep the
system arcing free the fluence, as a rule of thumb, should be
below 20~mJ/cm$^2$ for picoseconds pulses and in presence of
voltage in excess of few hundred kilovolts.

\begin{acknowledgments}
We wish to thanks E. Kirk and T. Garvey for careful proofreading
this manuscript.
\end{acknowledgments}

%
%

\end{document}